\newcommand{\nd}{Ni(C$_5$D$_{14}$N$_2$)$_2$N$_3$(PF$_6$)}
\newcommand{\nenp}{Ni(C$_2$H$_8$N$_2$)$_2$NO$_2$(ClO$_4$)}
\newcommand{\ndmaz}{Ni(C$_5$H$_{14}$N$_2$)$_2$N$_3$(ClO$_4$)}

\documentclass[aps,prl,twocolumn,superscriptaddress]{revtex4}
\usepackage{graphicx}

\begin{document}

\title{Quasi-elastic neutron scattering in the high-field phase of a Haldane antiferromagnet.}

\author{A. Zheludev}
\affiliation{Physics Department, Brookhaven National Laboratory,
Upton, NY 11973-5000, USA.}

\affiliation{Present address: Solid State Division, Oak Ridge
national Laboratory, Oak Ridge, TN  37831-6393, USA.}

\author{Z. Honda}
\affiliation{Faculty of Engineering, Saitama University, Urawa,
Saitama 338-8570, Japan.}

\author{Y. Chen}
\affiliation{Department of Physics and Astronomy, Johns Hopkins
University, Baltimore, MD 21218, USA.}

\author{C. L. Broholm}
\affiliation{Department of Physics and Astronomy, Johns Hopkins
University, Baltimore, MD 21218, USA.}

\affiliation{NIST Center for Neutron Research, National Institute
of Standards and Technology, Gaithersburg, MD 20899, USA.}

\author{K. Katsumata}
\affiliation{The RIKEN Harima Institute, Mikazuki, Sayo, Hyogo
679-5148, Japan.}

\author{S. M. Shapiro}
\affiliation{Physics Department, Brookhaven National Laboratory,
Upton, NY 11973-5000, USA.}

\date{\today}
\begin{abstract}
Inelastic neutron scattering experiments on the Haldane-gap
quantum antiferromagnet \nd\ are performed in magnetic fields
below and above the critical field $H_c$ at which the gap closes.
Quasi-elastic neutron scattering is found for $H>H_c$ indicating topological excitations in the high field phase.
\end{abstract}

\pacs{75.50.Ee,75.10.Jm,75.40.Gb}

\maketitle \narrowtext

\begin{figure}
\includegraphics[width=3.3in]{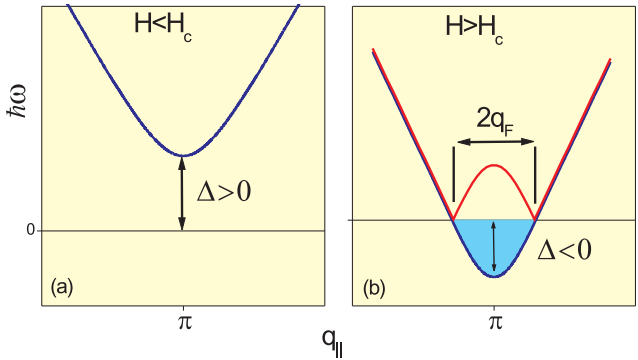}
   \caption{Dispersion in the lowest-energy Haldane-gap mode below (a)
and above (b) the critical field $H_c$ (blue). In the free-fermion
model, excitations with a negative energy at $H>H_c$ form a Fermi
sea (cyan) with a characteristic Fermi vector $q_{\rm F}$ Lowest
energy excited states are viewed as ``particles'' for
$|q_\|-\pi|>q_{\rm F}$ or ``holes'' for $|q_\|-\pi|<q_{\rm F}$
(red line). }\label{cartoon}
\end{figure}

As first realized by Haldane, integer-spin one-dimensional (1D)
Heisenberg antiferromagnets (AFs) are exotic ``quantum spin
liquids'' with only short-range spin correlations and a gap in the
magnetic excitation spectrum \cite{Haldane}. Haldane-gap systems
exhibit numerous unusual properties, and particularly interesting
predictions were made for their behavior in high magnetic fields
\cite{Schulz86,Affleck,Takahashi,Mitra94,Sachdev94,Konik01}. In an
external field the gap excitations, which are a $S=1$ triplet, are
subject to Zeeman splitting. The gap $\Delta$ for one of the three
branches decreases with field (Fig.~\ref{cartoon}(a)) and closes
at some critical field $H_c\sim \Delta/g\mu_{B}$
\cite{Schulz86,Affleck,Takahashi,Katsumata89}. Near $H_c$ this
problem is equivalent to Bose condensation
\cite{Affleck,Takahashi}. However, at $H>H_c$ the formation of a
condensate is prevented by strong interactions between magnons. As
a result, the high-field phase is rather unusual, with power-law
spin correlations, and is an example of a {\it Luttiner spin
liquid} \cite{Sachdev94,Konik01}. The basic physics at $H>H_c$ can
be understood within a simple model.
The magnon interactions are
assumed to be a hard-core repulsion, in which case the excitations
behave as free fermions \cite{Affleck,Takahashi,Tsvelik,Fujiwara}.
At $H>H_c$ the ground state is a Fermi sea of excitations with
field-dependent Fermi density and wave vector $q_F$, as
illustrated in Fig.~\ref{cartoon}(b). The spectrum is gapless, and
dynamic spin correlations are incommensurate. The effect is
related to field-induced incommensurability in gapless spin
chains, that is described using a different fermion mapping
\cite{S12}.

 For a long
time the spin dynamics of Haldane gap AFs in the high-field phase
evaded direct experimental investigation by neutron scattering
techniques. This was primarily due to large values of $H_c$ for
most known model quasi-1D integer-spin compounds.  The situation
changed with the discovery of \nd\ (NDMAP), a material well-suited
for neutron scattering experiments and a critical field of only
5~T\cite{Honda}. For $H>H_c$ residual interactions between the
$S=1$ Ni$^{2+}$-chains in this compound lead to commensurate
long-range AF ordering at low temperatures \cite{Honda,Chen01}.
However, the strength of effective inter-chain coupling is about
0.2~meV, small compared to the in-chain exchange constant
$J\approx 2.8$~meV \cite{Zheludev01,Koike00}. At
$\hbar\omega\gtrsim 0.2$~meV one can thus directly study the
exotic dynamics of the 1D model. In the present work we report
inelastic neutron scattering studies of low-energy excitations in
NDMAP, for magnetic field below and well above $H_c$, at
temperatures above the 3D-ordered phase.

\begin{figure}
\includegraphics[width=3.3in]{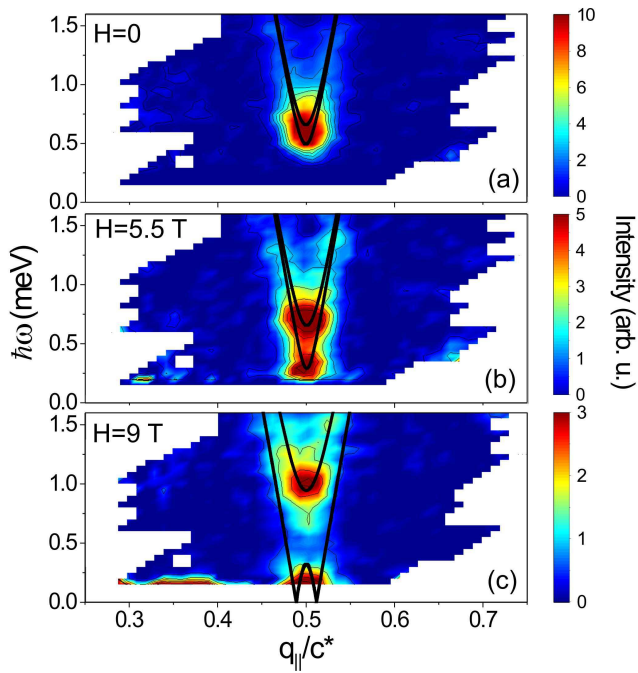}
 \caption{False color
plots of inelastic intensity measured in NDMAP at $T=2.5$~K for
different values of magnetic field applied along the $b$
axis. The extra intensity seen below
  $q_{\|}= 0.4c^{\ast}$ at low energies in (c) is most likely an artifact of imperfect background
  subtraction. Solid lines are as described in the text.}\label{exdata}
\end{figure}

Experiments were carried out on the SPINS 3-axis spectrometer
installed at the cold neutron facility at the National Institute
of Standards and Technology Center for Neutron Research. 23
deuterated NDMAP single crystals were co-aligned to produce a
sample of total mass 1.2 g and a mosaic of about 7$^{\circ}$. The
sample was mounted in a cryomagnet and measurements were performed
at $T=2.5$~K and magnetic fields $H=0$, 2, 4, 5.5, 7 and 9~T. The
field was applied along the $b$ axis of the orthorhombic structure
(space group $Pnmn$), perpendicular to the spin chains that run
along $c$ (cell parameters $a=18.05$~\AA, $b=8.71$~\AA,
$c=6.14$~\AA). The spectrometer employed a vertically focusing
pyrolitic graphite (PG) (002) monochromator and a 23~cm$\times
15$~cm flat multi-blade PG(002) analyzer positioned 91.4~cm from
the sample. Scattered neutrons were registered by a
position-sensitive detector (PSD). Measurements were done with
fixed-final neutron energy  for the central analyzer blade of
3.1~meV.  The collimation setup was
($^{58}$Ni-guide)--(open)--(open)--80'(radial), and a Be filter
was used either before or after the sample. To minimize
the contributions of the sample mosaic to wave vector
resolution, all scans were performed such that momentum transfer
for the center pixel of the PSD was along $c^*$. Wave vector transfer along $c^*$ will be denoted by $q_{\|}$. In a single setting of the
spectrometer the PSD covered an energy range of 1 meV and a range
of wave vector transfer along the chain of 0.3 $c^*$ Separate
scans were corrected for the analyzer efficiency, and merged to
produce a single data set. A linear background was subtracted from
each const-$E$ slice taken from the data set at each field. The
data were smoothed using a Gaussian kernel with $\delta E=0.05$~meV FWHM
and $\delta q_{\|}=0.01$ $c^*$. Typical resulting data sets are
visualized in Fig.~\ref{exdata}.

For the following discussion it is important to estimate the
critical field $H_c$. Long-range ordering, that is essentially a
three-dimensional (3D) effect and occurs at $H_c^{3D}\approx 12$~T
for $T=2.5~K$, is of little interest in the present study. The
relevant quantity is the field $\tilde{H}_c$ at which the gap for
the lower-energy Haldane excitation vanishes at
$\bbox{q}=(0,0,0.5)$, where all the measurements were performed.
$\tilde{H_c}$ can be estimated from the gap energies measured in
zero field\cite{Zheludev01} and the known gyromagnetic ratio for
the $S=1$-carrying Ni$^{2+}$-ions in NDMAP \cite{Honda}. Using
Eq.~2.14 in Ref.\cite{Golinelli93} we get $\tilde{H_c}\approx
7.3$~T. Below we shall first review the results obtained for
$H<\tilde{H_c}$, and proceed to discuss the 9~T data, that
represent the high-field phase and contain our most important
findings.

The Haldane triplet in NDMAP is split by single-ion anisotropy
\cite{Zheludev01}. The parabolic shape seen in
Fig.~\ref{exdata}(a) for $H=0$ corresponds to two lower-energy gap
excitations polarized along the $a$ and $b$ axes, with gap
energies $\Delta_1= 0.48(1)$~meV and $\Delta_2= 0.65(1)$~meV,
respectively. The experimental resolution is not sufficient to
resolve these two modes. The gap in the $c$-axis polarized Haldane
excitation, $\Delta_3\approx1.9$~meV \cite{Zheludev01}, is outside
the energy range covered in the present experiment. In
Fig.~\ref{exdata}(b), at $H=5.5$~T, one clearly sees the splitting
of the doublet. Excitations polarized along the $b$ axis (upper
branch) are $S_z=0$ states, $z$ being chosen along the applied
field. To first order, they are not affected by the magnetic
field. The gap for the $a$-axis polarized Haldane excitation
(lower mode) decreases with field by virtue of the Zeeman effect.

\begin{figure}
\includegraphics[width=3.3in]{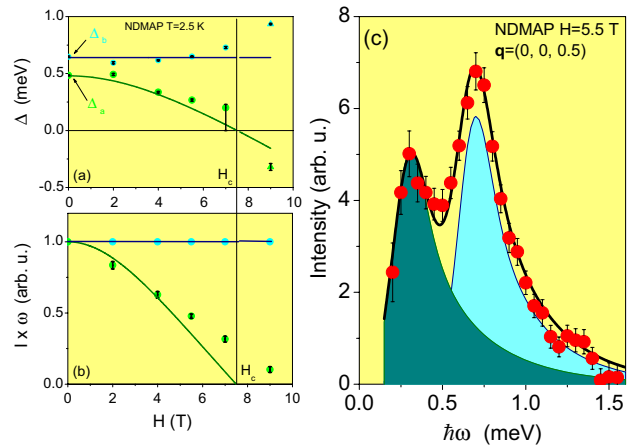}
 \caption{Field dependence of the gap energies (a) and excitation intensities (b) measured in NDMAP at $T=2.5$~K (symbols).
  (c) A constant-$Q$ scan collected  at $H=5.5$~T (symbols). Solid lines and shaded areas are
 as described in the text.}\label{fits}
\end{figure}

\begin{figure}
\includegraphics[width=3.3in]{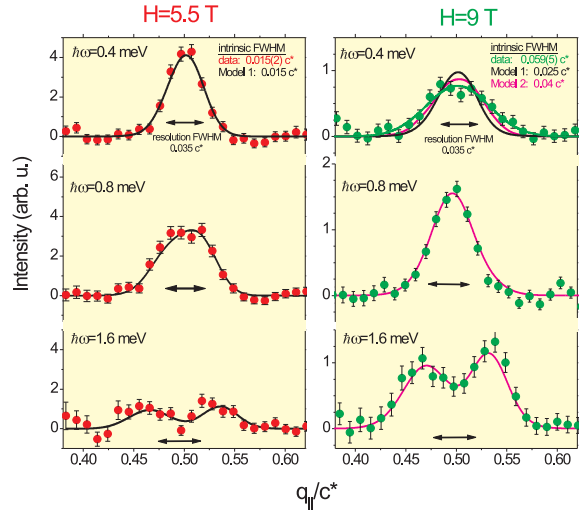}
 \caption{
 Symbols: constant-$E$ scans collected in NDMAP at $H=5.5$~T$<\tilde{H}_c$ (a--c) and $H=9$~T$>\tilde{H}_c$ (d--f).
 Lines are as described in the text. Arrows indicate the projection of the resolution ellipse
 onto the chain axis.}\label{conste}
\end{figure}

The data were analyzed using a parameterized model cross section,
numerically convoluted with the calculated spectrometer
resolution. Details of the data analysis will be given elsewhere
and only the main points are summarized here. The cross section
was written in the single mode approximation \cite{Zheludev01}.
For each observed excitation branch the dispersion relation was
written as
$(\hbar\omega_{\bbox{q}})^2=\Delta^2+v^2\sin^2(\bbox{q}\bbox{c})$
(Model 1). The gaps and intensities for each branch were adjusted
to best fit the experimental data. At each field the {\it entire}
data set with energy transfers between 0.2~meV and 1.6~meV was
analyzed with a single set of parameters in a {\it global} fit.
The spin wave velocity $v$ was fixed at $v=6.9$~meV, as determined
at $H=0$. The model was found to describe the data measured at up
to 7~T very well, with a residual $\chi^2<1.5$ in all cases. The
obtained field dependence of the gap energies is plotted in
Fig.~\ref{fits}(a). The corresponding dispersion relations are
shown in solid lines in Fig.~\ref{exdata}(a)-(b).
Fig.~\ref{fits}(c) shows a constant-$Q$ scan extracted from the
5.5~T data set (symbols) plotted over the profile simulated using
the refined parameter values (black line). The shaded areas show
partial contributions from the two gap modes. Similarly,
Fig.~\ref{conste}(a)-(c) shows a series of extracted constant-$E$
scans for $H=5.5$~T, plotted on top of the corresponding simulated
profiles (red lines). The measured field dependence of gaps
follows the general trend [solid lines in Fig.~\ref{fits}(a)]
expected from a perturbation treatment of the Zeeman term
\cite{Golinelli93}. In agreement with the perturbation analysis
[blue line in Fig.~\ref{fits}(b)], the measured intensity of the
upper mode is field independent [Fig.~\ref{fits}(b), cyan symbols]
while that of the lower mode decrease with $H$ as it acquires
polarization along the c-direction to which our experiment is
insensitive [Fig.~\ref{fits}(c), green line]. At $H<\tilde{H}_c$
the behavior is well described by first order perturbation theory
and is similar to that previously found in the extensively studied
Haldane gap compound \nenp~(NENP) \cite{Regnault94}.

We now turn to the data visualized in Fig.~\ref{exdata}(c),
collected at $H=9$~T$>\tilde{H}_c$. The gap in the upper mode has
increased substantially, to about 0.93~meV, as compared to
0.63~meV at $H=5.5$~T. Interestingly, below this gap, a weak but
well-defined area of scattering extends all the way down to the
lowest accessible energy transfer. Due to limited wave vector
resolution and the steep  $1/\omega$ intensity scaling, the
dispersion of these excitations is difficult to discern other than
by a quantitative analysis (see below). This is not conventional
critical scattering associated with the imminent phase transition,
as the experiment was done away from ${\bf Q}_c=(h,0.5,0.5)$ where
magnetic Bragg peaks appear below $T_N$. This first experimental
observation of quasi-one-dimensional quasi-elastic neutron
scattering for $H>H_c$ implies that individual spin chains in
NDMAP are in a critical phase from which quasi-long-range order
would likely emerge on cooling even in the absence of inter-chain
coupling.

Closer examination of the data reveals important changes at high fields beyond closing the gap in the neutron scattering spectrum.
Applying the low field analysis procedure to the 9 T data yields a significantly worse fit ($\chi^2=2.0$) than was achieved for $H<\tilde{H}_c$. The problem is mainly in the wave vector and
energy dependence of inelastic scattering at low energies. The
discrepancy is particularly pronounced in the shape of
constant-$E$ scans, where the data shows peaks considerably
broader than simulations based on parameters obtained in the fit.
To quantify this behavior we scaled the measured intensity by the
energy transfer and integrated the result over the range
0.25--0.5~meV. The resulting $q$-dependence was fit to a Gaussian
profile to obtain an intrinsic (resolution corrected) width of
$0.048(5)c^{\ast}$. The peak simulated using parameters determined
in the global fit has an intrinsic width of only $0.023c^{\ast}$.
The anomalous $q$-width is also visible in each measured
constant-$E$ scan individually, as for example at 0.4~meV energy
transfer [Fig.~\ref{conste} (d)]. It is important to emphasize
that the discrepancy is {\it not} related to an error in
resolution calculation. Indeed, at lower fields, {\it e. g.} at
$H=5.5$~T [Fig.~\ref{conste} (a)] Gaussian fits to const-$E$ scans
and simulated profiles yield identical widths. Moreover, even at
$H=9$~T at higher energy transfers, where the dominant
contribution is from the 0.93~meV mode, the model cross section
fits the data quite well [Fig.~\ref{conste} (e) and (f), solid
lines].

At least two theoretical frame works could potentially account for the $q_{\|}-$broadened quasi-elastic scattering in the high field phase of NDMAP. A classical easy plane antiferromagnet with a field in the easy plane can be mapped onto a Sine-Gordon model with a gapped spin wave spectrum and topological soliton excitations.\cite{mikeskasteiner} While solitons have a finite rest mass, they are stable topological objects that move freely along the spin chain. The soliton gas yields quasi-elastic neutron scattering with a $q_{\|}-$width that varies in proportion to the soliton density. The soliton model was used successfully to account for the phase diagram of NDMAP with fields normal to the chain axis.\cite{Honda01} In addition Soliton pinning at defects may account for the quasi-two-dimensional frozen state in NDMAP for fields perpendicular to the easy plane.\cite{Chen01} Further evaluation of the Sine-Gordon model for the high field phase of NDMAP would require measurements of the $H/T$ dependence of the $q_{\|}-$width for comparison to the theoretical soliton density. The free-fermion model presented in the introduction describes a different type of topological excitations. In this model at $H>H_{c}$ the lowest energy excitations with momentum transfers
along the chains near $q_\|=\pi/c$ are single-particle or
single-hole states. These have a dispersion relation as shown by
the red line in [Fig.~\ref{cartoon} (b)]. To compare the 9~T data to the free fermion model, we constructed an empirical dispersion relation for the lower mode [Model 2] with $q_F$ as a new parameter. The dispersion was assumed to be linear outside the
region $\pi/c-q_{\rm F}<q_\|< \pi/c+q_{\rm F}$, and to follow half
a period of a sine wave within these boundaries. The prefactor for
the sinusoidal part was chosen to give the correct spin wave
velocity at $q_\| = \pi/c\pm q_{\rm F}$. As illustrated in
Fig.~\ref{cartoon} (b), a given value of $q_F$ corresponds to a
``negative gap'' $\Delta=-{2 q_F cv/ \pi}$. With $q_{\rm
F}=0.03(1)c^{\ast}$ ($\Delta=-0.3(1)$~meV) Model 2 yields a
considerably better agreement with experiment ($\chi^2=1.3$) than
Model 1. The refined gap energy for the upper mode and the
``negative gap'' for the lower branch are indicated with triangles
in Fig.~\ref{fits} (a). Simulations based on resulting parameter
values are shown in magenta lines Fig.~\ref{conste} (d)-(f), and the dispersion
relations are shown in a solid line in Fig.~\ref{exdata}(c). While
at high energies where the upper mode is dominant the two Models are
virtually indistinguishable, below 0.6~meV transfer Model 2 does a better job at reproducing the observed peak
widths [Fig.~\ref{conste} (d)].

The free fermion model makes a very robust prediction for the
field dependence of $q_F$. The Fermi vector $q_F$ determines the
number of $S_z=1$-carrying particles present in the ground state
at $H>H_c$ and is therefore directly proportional to the uniform
magnetization: $M=g \mu_{\rm B}~ 2 q_F /c^{\ast}$, where $M$ is
the magnetization per spin. For NDMAP the magnetization curve has
been measured experimentally \cite{Honda00PhysicaB} and at 9~T
$M\approx 0.1 \mu_{\rm B}/$Ni$^{2+}$. This gives $ q_F=0.025
c^{\ast}$, remarkably close to the experimental value. It is
important to stress that the free-fermion model is a rather crude
approximation. In fact, at $H>H_c$ the system is to be described
as a sea of {\it interacting} fermions, {\it i. e.} a {\it
Luttinger liquid} (for a recent discussion see
Ref.~\cite{Konik01}). As a result of these interactions, there is
no well-defined Fermi surface or even single-particle poles in the
dynamic susceptibility. Incommensurate correlations are obscured:
the equal-time spin correlation function is a broad peak at the
commensurate $q_\|=\pi$ point. It is probably for this reason that
{\it static} long-range ordering in NDMAP\cite{Chen01} and the
related compound \ndmaz\ (NDMAZ) \cite{Zheludev01Z} is
commensurate. In the Luttinger liquid model the low energy
excitations are a continuum. Experimental evidence for this
behavior was recently obtained in studies of NDMAP\cite{Honda01}
and NENP \cite{ZalNENP}. While the resolution of our experiments on
NDMAP is insufficient to distinguish between continuum and
single-particle excitations, the free-fermion model appears to be
a good starting point for the data analysis. It can reproduce the
measured data quite well, and yields a {\it self-consistent}
estimate for $q_{\rm F}$.

In summary, our experiments demonstrate $q_{\|}-$broadened
quasi-elastic scattering in the high-field phase of a $S=1$
Haldane-gap compound. Various models based on topological
excitations with low energy phase fluctuations may account for the
data, including the Sine-Gordon and the free fermion models.
Further theoretical and experimental work will be required to
identify the correct description of the high field phase in the
anisotropic spin-1 chain.

We would like to thank I. Zaliznyak, A. Tsvelik, I. Affleck, and
S. Sachdev for illuminating communications, and R. Rothe (BNL) for
technical assistance. Work at BNL and ORNL was carried out under
DOE Contracts No. DE-AC02-98CH10886 and DE-AC05-00OR22725,
respectively. Work at JHU was supported by the NSF through DMR
9801742. Experiments at NIST were supported by the NSF through
DMR-9413101. Work at RIKEN was supported in part by a Grant-in-Aid
for Scientific Research from the Japanese Ministry of Education,
Culture, Sports, Science and Technology.


\end{document}